\documentstyle[12pt]{article}
\newtheorem{theorem}{Theorem}
\begin{document}
\title{
 Free Energy of Gravitating Fermions
 }
\author{
Neven Bili\'c$^1$
and
Raoul D.~Viollier$^2$
 \\
 $^1$Rudjer Bo\v{s}kovi\'{c} Institute, 10000 Zagreb, Croatia \\
 {\normalsize  bilic@thphys.irb.hr ; fax: 385-1-428 541} \\
$^2$Department of Physics,
University of Cape Town \\ Rondebosch 7700, South Africa
}
\date{\today}
\maketitle
\begin{abstract}
We study
 a system of self-gravitating
 massive fermions
in the framework
of the general-relativistic Thomas-Fermi model.
We postulate the free energy functional
and show that its extremization
is equivalent to solving the
Einstein's field equations.
A self-gravitating fermion gas
we then describe by
a set of Thomas-Fermi type  self-consistency
equations.
\end{abstract}
%

%

 Thermodynamical properties of the self-gravitating
 fermion gas have been extensively studied
 in the framework of the Thomas-Fermi model
 [1-6].
 The system was investigated in the nonrelativistic
 Newtonian limit.
   The canonical and grand-canonical ensembles
     for such a system
  have been shown to
have a nontrivial thermodynamical limit~\cite{thi,her1}.
Under certain conditions this system
 will undergo a phase transition
 that is accompanied by a gravitational collapse~\cite{her1,mes}
 which may have important astrophysical and cosmological
 implications~\cite{bil}.

 In this paper we formulate the general-relativistic
 version of the model.
 The effects of general relativity become important
 if the total rest-mass of the system is close to the
 Oppenheimer-Volkoff limit \cite{opp}.
 There are three main features that distinguish the relativistic
 Thomas-Fermi theory
  from the Newtonian one:
  {\it i}) the equation of state is relativistic
  {\it ii}) the temperature and chemical potential are
  metric dependent local quantities
  {\it iii}) the gravitational potential satisfies Einstein's field
  equations (instead of Poisson's equation).

Let us first discuss the general properties
of a canonical, self-gravitating relativistic fluid.
Consider a nonrotating fluid consisting of $N$ particles
in
a spherical volume of radius $R$ in equilibrium
at non-zero temperature.
We denote by
 $u_{\mu}$ ,
$p$, $\rho$, $n$ and $\sigma$ the velocity, pressure,
energy density, particle number density and
entropy density of the fluid.
Following Gibbons and Hawking \cite{gib} we postulate
quite generally
the free energy of the canonical ensemble as
\begin{equation}
F=M-\int_{\Sigma} T\sigma \, k^{\mu}d\Sigma_{\mu} \, ,
\label{eq30}
\end{equation}
where $M$ is the total mass as measured from infinity,
$\Sigma$ is a spacelike hypersurface that contains
the fluid and is orthogonal to the time-translation
Killing vector field $k^{\mu}$
which is related to the velocity of the fluid
\begin{equation}
k^{\mu}=\xi u^{\mu}\, ;  \;\;\;\;\;\;
\xi=(k^{\mu}k_{\mu})^{1/2}.
\label{eq50}
\end{equation}
 The entropy density of a relativistic fluid may be expressed as
\begin{equation}
\sigma=\frac{1}{T}(p+\rho-\mu n),
\label{eq16}
\end{equation}
and
 $T$  and
$\mu$ are
the local temperature and chemical potential.
The metric generated by the mass distribution
is  static, spherically symmetric and asymptotically
flat, i.e.
\begin{equation}
ds^2=\xi^2 dt^2 -\lambda^2 dr^2 -
     r^2(d\theta^2+\sin \theta d\phi^2).
\label{eq00}
\end{equation}
$\xi$ and $\lambda$ may be represented in terms of the
gravitational potential and mass
\begin{equation}
\xi=e^{\varphi (r)},
\label{eq01}
\end{equation}
\begin{equation}
\lambda=\left(1-\frac{2m(r)}{r}\right)^{-1/2}
\label{eq10}
\end{equation}
with
\begin{equation}
m(r)=\int^r_0 dr'\, 4\pi r'^2  \rho(r')   \, .
\label{eq11}
\end{equation}
\\
The equation of hydrostatic equilibrium
\cite{tol,lan}
\begin{equation}
\partial_{\nu}p=-(p+\rho)\xi^{-1}\partial_{\nu}\xi ,
\label{eq17}
\end{equation}
together with the thermodynamic identity
\begin{equation}
 d\frac{\mu}{T}=\frac{1}{n}d\frac{p}{T}+\frac{\rho}{n}d\frac{1}{T},
\label{eq18}
\end{equation}
implies
\begin{equation}
T \xi=T_0\, ; \;\;\;\;\;\;
\mu \xi=\mu_0  \, ,
\label{eq21}
\end{equation}
where $T_0$ and $\mu_0$ are constants equal to the
temperature and chemical potential at infinity.
$T_0$ may be chosen arbitrarily, but $\mu_0$ is
an implicit functional of $\xi$ due to
the constraint that the total number of particles
\begin{equation}
\int_{\Sigma} n\, u^{\mu}d\Sigma_{\mu}
=N
\label{eq26}
\end{equation}
is fixed.

Based on Eq. (\ref{eq21}) the free energy may be written
in the form analogous to ordinary thermodynamics
\begin{equation}
F=M-T_0 S
\label{eq60}
\end{equation}
with  $M=m(R)$
and the total entropy $S$ defined as
\begin{equation}
S = \int_0^R dr\,4\pi r^2 \lambda
\frac{1}{T}(p+\rho)-\frac{\mu_0}{T_0} N  ,
\label{eq70}
\end{equation}
where we have employed the spherical symmetry to
replace the proper volume integral as
\begin{equation}
\int_{\Sigma} u^{\mu}d\Sigma_{\mu}
= \int_0^R dr 4\pi r^2 \lambda .
\label{eq80}
\end{equation}

The following theorem shows
that the quantity defined by
  Eq. (\ref{eq30})
is indeed the free energy of a self-gravitating
relativistic fluid at finite temperature.

\begin{theorem}
Among all momentarily
static, spherically symmetric configurations
$\{\xi(r),m(r)\}$
which for given temperature $T_0$ at infinity
 contain a specified number of particles
\begin{equation}
 \int_0^R 4\pi r^2 dr \, \lambda(r)  n(r) = N
\label{eq25}
\end{equation}
within a spherical volume of given radius
 $R$,
those and only those  configurations
that
extremize the quantity F defined by
{\rm (\ref{eq60})}
 will
 satisfy the Einstein's field equation
\begin{equation}
\label{eq22}
\frac{d\xi}{dr}=\xi\frac{m+4\pi r^3 p}{r(r-2m)} \, ,
\end{equation}
with the boundary condition
\begin{equation}
\xi(R)=\left(1-\frac{2 M}{R}\right)^{1/2}.
\label{eq23}
\end{equation}
\end{theorem}
{\it Proof.}
From Eqs. (\ref{eq60}) and (\ref{eq70})
by making use of the identity (\ref{eq18}),
 and the fact
that $\delta(\mu/T)=\delta(\mu_0/T_0)$
and that $N$ is fixed by the constraint (\ref{eq25}), we find
\begin{equation}
\delta F= \delta M -
\int_0^R dr\, 4\pi r^2 \frac{T_0}{T}(p+\rho)
\delta \lambda
-  \int_0^R dr\, 4\pi r^2 \lambda \frac{T_0}{T} \delta\rho \, .
\label{eq90}
\end{equation}
The variations $\delta\lambda$ and $\delta\rho$
 can be expressed in terms of the variation
$\delta m(r)$
and its derivative
\begin{equation}
\frac{d\delta m}{dr} =4\pi r^2 \delta\rho.
\label{eq93}
\end{equation}
This  gives
\begin{equation}
\delta F= \delta M -
\int_0^R dr\, 4\pi r^2
 \frac{T_0}{T}(p+\rho)
\frac{\partial\lambda}{\partial m}
\delta m
-\int_0^R dr\, \lambda\frac{T_0}{T}\frac{d\delta m}{dr}.
\label{eq91}
\end{equation}
By partial integration of the last term
and replacing $T_0/T$ by $\xi$ we find
\begin{equation}
\delta F =
\left[1-\lambda(R)\xi(R)\right]\delta M
- \int_0^R dr\, \left[4\pi r^2 \xi (p+\rho)
\frac{\partial\lambda}{\partial m}
-\frac{d}{dr}(\lambda\xi)\right]\delta m \, ,
\label{eq94}
\end{equation}
where $\delta m(r)$ is an arbitrary variation
on the interval $[0,R]$,
except for
the constraint
$\delta m(0)=0$.
Therefore $\delta F$ will vanish if and only if
\begin{equation}
4\pi r^2 \xi (p+\rho)
\frac{\partial\lambda}{\partial m}
-\frac{d}{dr}(\lambda\xi) =0
\label{eq95}
\end{equation}
and
\begin{equation}
1-\lambda(R)\xi(R) =0.
\label{eq96}
\end{equation}
Using Eqs. (\ref{eq10}) and (\ref{eq11}) we can write
Eq. (\ref{eq95}) in the form (\ref{eq22}),
and Eq. (\ref{eq96}) gives the desired boundary condition
 (\ref{eq23}).
Thus,
$\delta F=0$ if and only if a configuration
$\{\xi,m\}$ satisfies Eq. (\ref{eq22})
with (\ref{eq23})
as was to be shown.
\\
{\it Remark 1.}
A solutions to Eq. (\ref{eq22})
is dynamically stable if
the free energy assumes a minimum.
\\
{\it Remark 2.}
Our Theorem 1 is a finite temperature generalization
of the result obtained for
cold, catalyzed matter \cite{har}.

We now proceed to the formulation of the general-relativistic
Thomas-Fermi model.
Consider the case of a self-gravitating
gas consisting of $N$ fermions with the mass
$m_f$ contained in a sphere of radius $R$.
 Given the temperature at infinity  $T_0$,
the following set of self-consistency
equations  defines
the  Thomas-Fermi equation:
\\
{\bf 1 Equation of state}
\begin{equation}
n   = g \int^{\infty}_{0} \frac{d^3q}{(2\pi)^3}\,
\frac{1}{1+e^{E/T-\mu/T}} \, ,
\label{eq13}
\end{equation}
\begin{equation}
\rho = g \int^{\infty}_{0} \frac{d^3q}{(2\pi)^3}\,
\frac{E}{1+e^{E/T-\mu/T}} \, ,
\label{eq14}
\end{equation}
\begin{equation}
p = g T \int^{\infty}_{0} \frac{d^3q}{(2\pi)^3}\,
\ln (1+e^{-E/T+\mu/T}) \, ,
\label{eq15}
\end{equation}
where
$g$ denotes the spin degeneracy factor,
$T$ and $\mu$ are local temperature and chemical potential,
respectively,
as defined in Eq. (\ref{eq21}),
 and $E=\sqrt{m_f^2+q^2}$.
\\
{\bf 2
 Field equations}
\begin{equation}
\frac{d\xi}{dr}=\xi\frac{m+4\pi r^3 p}{r(r-2m)} \, ,
\label{eq42}
\end{equation}
\begin{equation}
\frac{dm}{dr}=4\pi r^2 \rho,
\label{eq43}
\end{equation}
with the boundary conditions
\begin{equation}
\xi(R)=\left(1-\frac{2 m(R)}{R}\right)^{1/2}
\, ; \;\;\;\;\;
m(0)=0.
\label{eq44}
\end{equation}
{\bf 3 Particle number constraint}
\begin{equation}
\int_0^Rdr\, 4\pi r^2 (1-2m/r)^{-1/2}\, n(r)=N .
\label{eq45}
\end{equation}
One additional important requirement is that a solution
of the self-consistency equations (\ref{eq13}) to (\ref{eq45})
have to minimize
the free energy
defined by Eq. (\ref{eq60})

Next we  show  that, in the
Newtonian limit, we recover the
usual Thomas-Fermi equation.
Using
the nonrelativistic chemical potential
$\mu_{NR}=\mu_0-m_f$
and
 the approximation
$\xi=e^{\varphi}\simeq 1+\varphi$,
$E\simeq m_f+q^2/2m_f$
and
  $m/r \ll 1$ ,
 we find the usual Thomas-Fermi self-consistency
 equations \cite{mes,bil}
\begin{equation}
n=\frac{\rho}{m_f}
 = g \int^{\infty}_{0} \frac{d^3q}{(2\pi)^3}\,
\left(1+\exp(\frac{q^2}{2m_fT_0}+\frac{m_f}{T_0}\varphi
 -\frac{\mu_{NR}}{T_0}) \right)^{-1} \, ,
\label{eq49}
\end{equation}
\begin{equation}
\frac{d\varphi}{dr}=\frac{m}{r^2} \, ;
\;\;\;\;
\frac{dm}{dr}=4\pi r^2 \rho  \,  ,
\label{eq41}
\end{equation}
\begin{equation}
\varphi(R)=-\frac{m_f N}{R}
\, ; \;\;\;
m(0)=0,
\label{eq47}
\end{equation}
\begin{equation}
\int_0^R dr\,4\pi r^2 n(r)=N.
\label{eq46}
\end{equation}
The free energy  (\ref{eq60}) in the Newtonian limit yields
\begin{equation}
F=m_f N +\mu_{NR} N - \frac{1}{2}\int_0^R dr \, 4\pi r^2 n\varphi
-\int_0^R dr \, 4\pi r^2 p
\label{eq40}
\end{equation}
with
\begin{equation}
 p= g T_0\int^{\infty}_{0} \frac{d^3q}{(2\pi)^3}\,
\ln\left(1+\exp(-\frac{q^2}{2m_fT_0}-\frac{m_f}{T_0}\varphi
 +\frac{\mu_{NR}}{T_0}) \right) \, ,
\label{eq48}
\end{equation}
which, up to a constant, equals the Thomas-Fermi free
energy \cite{her2}.

A straightforward thermodynamic limit
$N\rightarrow\infty$
as discussed by
Hertel, Thirring and Narnhofer \cite{her1,her2}
is in our case not directly applicable.
First, in contrast to the non-relativistic case,
 there exist, at zero temperature, a limiting configuration
with maximal $M$ and $N$ (Oppenheimer-Volkoff limit)
and something similar can be expected at
$T\neq 0$.
Second,
the scaling properties of the relativistic
Thomas-Fermi equation are quite distinct from
the nonrelativistic one.
The following scaling property can be easily shown:
  \\
  If the configuration
 $\{\xi(r),m(r)\}$ is a solution to the self consistency
 equations (\ref{eq13}) to (\ref{eq45}), then the
 configuration
 $\{\tilde{\xi}=\xi(A^{-1}r),\tilde{m}=Am(A^{-1}r);A>0\}$
 is also a solution with
 the rescaled
 fermion number
 $\tilde{N}=A^{3/2}N$,
 radius
 $\tilde{R}=AR$,
 asymptotic temperature
 $\tilde{T_0}=A^{-1/2}T_0$,
 and fermion mass
 $\tilde{m_f}=A^{-1/2}m_f$.
 The free energy is then rescaled as
 $\tilde{F}=AF$.
 Therefore, there exist a thermodynamic limit
 of
 $N^{-2/3}F$,
 with
 $N^{-2/3}R$,
 $N^{1/3}T_0$,
 $N^{1/3}m_f$
 approaching constant values
  when
 $N\rightarrow\infty$.
 In that limit
 the Thomas-Fermi equation becomes exact.
 It is obvious that application of this model to stellar systems
 should work very well if the interactions
 among individual particles are negligible.

\vspace{0.2in}
\newpage
%
\end{document}